\documentclass[preprint,aip,showpacs]{revtex4}

\usepackage{graphicx}

\usepackage{dcolumn} 

\usepackage{color}

\usepackage{bm}      

\usepackage{tabularx}

\usepackage{float} 

\newcommand{\vk}{\bf k}

\newcommand{\kpar}{k_{\parallel}}

\begin{document}

\title{Hall effect in a strong magnetic field: Direct comparisons of compressible magnetohydrodynamics and the reduced Hall magnetohydrodynamic equations.}

\author{L.N. Martin$^1$, P. Dmitruk$^{1}$, D. O. Gomez$^{1,2}$}

\affiliation{$^1$ Departamento de F\'\i sica, Facultad de Ciencias Exactas y
         Naturales, Universidad de Buenos Aires and IFIBA, CONICET, Ciudad 
         Universitaria, 1428 Buenos Aires, Argentina. \\
             $^2$ Instituto de Astronomia y Fisica del Espacio, Ciudad Universitaria, 1428 Buenos Aires, Argentina.}

\date{\today}

\begin{abstract}

In this work we numerically test a model of Hall magnetohydrodynamics 
in the presence of a strong mean magnetic field, the reduced Hall MHD model (RHMHD) 
derived by Gomez et al. \cite{D2}, with the addition of weak 
compressible effects. The main advantage of this model lies in the reduction of computational cost. 
Nevertheless, up until now the degree of agreement with the original 
Hall MHD system and the range of validity in a regime of turbulence
was not established. In this work direct numerical simulations of three-dimensional Hall 
MHD turbulence in the presence of a strong mean magnetic field are compared 
with simulations of the weak compressible RHMHD model. 
The results show that the degree of agreement is very high 
(when the different assumptions of RHMHD, like spectral anisotropy,
are satisfied). Nevertheless, when 
the initial conditions are isotropic but 
the mean magnetic field is maintained strong, the results differ 
at the beginning but asymptotically reach a good agreement at relatively 
short times. We also found evidence that the compressibility still plays 
a role in the dynamics of these systems, and the weak compressible 
RHMHD model is able to capture these effects. In conclusion the weak 
compressible RHMHD model is a valid approximation of the Hall MHD turbulence 
in the relevant physical context.

\end{abstract}

\maketitle

\section{Introduction\label{intro}}

The magnetohydrodynamic (one-fluid) model is often regarded as a 
reasonable description of the dynamics of a plasma. One-fluid models are useful
in the context of large scale dynamics, but when a more detailed description 
is needed (for instance, when the physical context favors the development of small scales) 
it is most appropriate to consider two-fluid models. At this level, the Hall
effect, which takes into account the separation between electrons and ions,
becomes relevant. The Hall length scale distinguishes the range of scales at 
which the Hall effect needs to be taken into account. When the Hall scale
is large enough to be separated from the dissipation length scale, an
approximation known as Hall MHD can be used to describe the dynamics.

The Hall MHD approximation consider two-fluid effects through a generalized Ohm's law which includes the Hall current.
Whenever one deals with phenomena with characteristic length scales comparable or smaller than the ion skin depth $c/\omega_i$
($c$: speed of light; $\omega_i$: ion plasma frequency), the Hall effect cannot be neglected. Among its other manifestations, the Hall
current causes (in an ideal plasma) the magnetic field to become frozen in the electron flow instead of being carried
along with the bulk velocity field. Another important feature of the ideal Hall-MHD description is the self-consistent presence
of parallel (i.e., to the magnetic field) electric fields. 
Hall-MHD has recently been
invoked in advancing our understanding of phenomena ranging from dynamo mechanisms \cite{i0}, magnetic reconnection\cite{i1,i2,i3}, and
accretion disks\cite{i4,i5} to the physics of turbulent regimes \cite{i6,i7,i8,i9}.

In many cases of interest, such as in fusion devices or geophysical and astrophysical plasmas, a strong externally supported magnetic field is present. This external field breaks the isotropy of the problem and can be responsible of important changes in the dynamics, such 
as in reconnection regimes or in the development of energy cascades in turbulent systems. For one-fluid MHD, the existence of a strong magnetic field is often exploited to yield a simpler model: the so-called reduced MHD approximation (RMHD; see refs. \cite{rhmhd_2} and \cite{montgomery} ). In this approximation, the fast compressional Alfv\'en mode is eliminated, while the shear Alfv\'en and the slow magnetosonic modes are retained \cite{i11}.
The RMHD equations have been used to investigate a variety of problem such as current sheet formation \cite{i12,i13}, nonstationary 
reconnection \cite{i14,i15}, the dynamics of coronal loops \cite{i16,i17}, or the development of turbulence \cite{i18}. The self-consistency of the RMHD approximation has been analyzed in ref. \cite{i19}. Moreover, recent numerical simulations have studied the validity of the RMHD equations by directly comparing its predictions with the compressible MHD equations in a turbulent regime \cite{i20}.

In this paper we test a new model derived in a recent work by Gomez et al. \cite{D2}. This model consists of a system of reduced
Hall-MHD (RHMHD) equations derived from the incompressible Hall MHD following the same asymptotic procedure, which is employed to obtain 
the conventional RMHD from MHD. The resulting set will describe the slow dynamics of a plasma (with Hall currents) embedded in a strong
external magnetic field and will naturally include new features such as the presence of a parallel electric field. Also we have made minor modifications to the model to describe the compressibility effects \cite{D3} and we tested its performance.

There are a number of applications for which it is reasonable to consider that the plasma is embedded in a 
strong and uniform magnetic field. Therefore, it is important to find new tools and methods 
to characterize systems where both the Hall effect and a strong mean magnetic
field are relevant. One of such tools are the direct numerical simulations. 
The full MHD models are computationally demanding and this demand is 
increased when the Hall effect is considered (because of numerical stability issues). 
On the other hand, the presence of a strong mean magnetic
field may simplify part of the dynamics. Our goal here is to test 
the RHMHD model \cite{D2}, which is aimed at reducing the computational cost.

The organization of the paper is as follows: Section \ref{modelos} describes the sets of equations for both compressible MHD and reduced 
Hall MHD, and the codes to numerically integrate these equations. In Section \ref{resultados} we present the numerical results: first a comparison between incompressible reduced and weakly compressible reduced MHD, second we study the solutions for different types of initial conditions. Finally, in Section \ref{conclusiones} we list our conclusions.

\section{Models and equations\label{modelos}}

\subsection{General model}

The compressible Hall MHD equations (dimensionless version) are

\begin{eqnarray} \label{n-s}
\frac{\partial{\boldsymbol{V}}}{\partial t}= \boldsymbol{V} \times \boldsymbol{\omega} + \frac{1}{M_{A}^{2}} \frac{\boldsymbol{J} \times \boldsymbol{B}}{\rho} - \nabla \left(\frac{\boldsymbol{V}^2}{2} + \frac{\rho^{\gamma-1}}{M_{S}^2(\gamma-1)} \right)  \nonumber \\ + \nu \frac{\nabla^{2}{\boldsymbol{V}}}{\rho} + (\zeta+\frac{1}{3}\nu) \frac{\nabla ({\nabla \cdot \boldsymbol{V}})}{\rho},
\end{eqnarray}

\begin{equation} \label{continuidad}
\frac{\partial{\rho}}{\partial t}=-\nabla \cdot (\rho \boldsymbol{V}),
\end{equation}

\begin{equation} \label{M1}
\frac{\partial{\boldsymbol{A}}}{\partial t}={\boldsymbol{V}}\times \boldsymbol{B} -\epsilon \frac{\boldsymbol{J} \times \boldsymbol{B}}{\rho} -\nabla \phi +\eta \nabla^2\boldsymbol{A},
\end{equation}

\begin{equation} \label{M2}
\nabla \cdot \boldsymbol{A}=0
\end{equation}

where $\boldsymbol{V}$ is the velocity field, $\boldsymbol{\omega}$ is the vorticity, $\boldsymbol{J}$ is the current, $\boldsymbol{B}$ the magnetic field, $\rho$ is the density of plasma, $\boldsymbol{A}$ and $\phi$ are the magnetic and electric potential. As shown in Eq. (\ref{n-s}) we used the barotropic law for the fluid, $p=cte.\rho^{\gamma}$ (here $p$ is the pressure), in our case we used $\gamma=5/3$. $M_S$ is the Mach number, $M_A$ is the Alfven number,  $\nu$ and $\zeta$ are the viscosities , $\eta$ is the resistivity and $\epsilon$ the Hall coefficient. All these numbers are control parameters in the numerical simulations.

We refer to this set of equations as the CMHD equations (Hall compressible MHD).

\subsection{Reduced models}

The RHMHD model derived in the work by Gomez $et$ $al$. \cite{D2} is a description of the two-fluid plasma dynamics in a strong external magnetic field. The model assumes that the normalized magnetic field is of the form (the external field is along $\boldsymbol{\widehat{e}_z}$)

\begin{equation} \label{1}
\boldsymbol{B}= \boldsymbol{\widehat{e}_z}+\delta\boldsymbol{B}, \mid\delta\boldsymbol{B}\mid\approx\alpha\ll1,
\end{equation}

where $\alpha$ represents the typical tilt of magnetic field lines with respect to the $\boldsymbol{\widehat{e}_z}$ direction, thus one expects

\begin{equation} \label{2}
\nabla_{\perp}\approx 1, \partial_z\approx\alpha\ll1.
\end{equation}

To ensure that the magnetic field $\boldsymbol{B}$ remains divergence free, it is assumed that

\begin{equation} \label{B}
\boldsymbol{B}= \boldsymbol{\widehat{e}_z} + \nabla \times(a \boldsymbol{\widehat{e}_z} + g \boldsymbol{\widehat{e}_x)}.
\end{equation}

For the same reason, a solenoidal (i.e. incompressible flow) velocity vector $\boldsymbol{V}$ is proposed,

\begin{equation} \label{V}
\boldsymbol{V}=\nabla \times(\varphi \boldsymbol{\widehat{e}_z} + f \boldsymbol{\widehat{e}_x)}.
\end{equation}

It is considered that the potentials $a(\boldsymbol{r},t)$, $g(\boldsymbol{r},t)$, $\varphi(\boldsymbol{r},t)$, and $f(\boldsymbol{r},t)$ are all of order $\alpha\ll1$. Introducing the expressions (\ref{B}) and (\ref{V}) in the compressible set of equations and taking the terms up to first and second order in $\alpha$ the RHMHD model is obtained.

In a similar fashion, a weakly compressible model can be obtained as well. To this end, a slight modification must be introduced to the hypothesis recently described \cite{D3}. The velocity field, in the more general case, can be decomposed as a superposition of a solenoidal part (incompressible flow) plus the gradient of a scalar field (irrotational flow), i.e.

\begin{equation} \label{V2}
\boldsymbol{V}=\nabla \times(\varphi \boldsymbol{\widehat{e}_z} + f \boldsymbol{\widehat{e}_x)} + \nabla \psi,
\end{equation}

Using the same assumptions as in \cite{D2}, but in addition assuming that $\psi$ is of order $\alpha^2$ the weak compressible RHMHD equations can be obtained,

\begin{equation} \label{rhmhd1}
\frac{\partial \omega}{\partial t}= B_0\frac{\partial j}{\partial x} + [j,a] - [\omega,\varphi] + \nu \nabla^2 \omega,
\end{equation}

\begin{equation} \label{rhmhd2}
\frac{\partial a}{\partial t}= B_0\frac{\partial (\varphi-\epsilon b)}{\partial x} + [\varphi,a] + [b,a] + \eta \nabla^2 a,
\end{equation}

\begin{eqnarray} \label{rhmhd3}
\frac{\partial b}{\partial t}= B_0\beta_p\frac{\partial (u-\epsilon j)}{\partial x} + [\varphi,b] + \beta_p[u,a]+\nonumber\\
 - \epsilon \beta_p[j,a] + \beta_p\eta\nabla^2b,
\end{eqnarray}

\begin{equation} \label{rhmhd4}
\frac{\partial u}{\partial t}= B_0\frac{\partial b}{\partial x} + [\varphi,u] - [a,b] + \nu \nabla^2 u,
\end{equation}

where
     
\begin{eqnarray}
\omega=-\nabla^2_\perp \varphi\label{omega},\\
j=-\nabla^2_\perp a\label{j},\\
b=-\partial_y g\label{b},\\
u=-\partial_y f\label{u},
\end{eqnarray}

and the notation $[A,B]=\partial_y A \partial_z B - \partial_y B \partial_z A$ is used.

The only difference between the weakly compressible RHMHD and the original RHMHD is in equation (\ref{rhmhd3}). In the original RHMHD equations $\beta_p=1$, whereas here $\beta_p=\beta\gamma/(1+\beta\gamma)$, with the plasma $\beta=4\pi p/B_0^2$.  
Therefore the compressibility effect here depends on the external magnetic field which embeds the plasma. Note that the entire flow compressibility is introduced through this parameter, which is a constant of the model. Therefore the computational cost is exactly the same as in the original RHMHD model.

We use this set of equations to study the dynamics of global magnitudes and compare them against the results from the full Hall MHD equations and its dependence on the initial conditions.

\subsection{Numerical codes\label{codigos}}

We use different pseudospectral codes to solve each set of equations.
Periodic boundary conditions are assumed in all directions of a cube of side $2\pi L$ (where $L\sim 1$ is the initial correlation
length of the fluctuations, defined as the length unit).
In the codes, Fourier components of the fluctuations are evolved in
time, starting from a specified set of Fourier modes (see following
section for the different types of initial conditions), with given total
energy and random phases. The RHMHD code specifies exactly the same 
initial fluctuation fields (velocity and magnetic field) as prepared
for the full MHD code. Initially, parallel component fluctuations are
set to zero.

The same resolution is used in both codes in all three directions in most of
the runs presented here, which have a moderate resolution of $128^3$ Fourier
modes, allowing many different runs to be done with different initial conditions
and/or mean magnetic field. The kinetic and magnetic Reynolds numbers
are defined as $R=1/\nu$, $R_m=1/\eta$, based on unit initial rms 
velocity fluctuation, unit length and non-dimensional values for the viscosity and
diffusivity. Here we considered $R=R_m=400$ ($\nu=\zeta=1/400$, $\eta=1/400$) in all of the runs. 
We also considered a Mach number $M_S=1/4$, Alfven number $M_A=1$, and Hall coeficient $\epsilon=1/16$
in all of the CMHD runs.

A second-order Runge-Kutta time integration is performed, the nonlinear
terms are evaluated using the standard pseudospectral procedure \cite{seudo}.
The runs are freely evolved for $10$ time units (the initial eddy turnover
time is defined in terms of the initial rms velocity fluctuation and unit
length). Two different values of the magnetic field $B_0 = 1, 8$ are considered (in units of the initial rms
magnetic fluctuation value).

\section{Results and discussion\label{resultados}}

We performed simulations with two different initial conditions, one of them maintains the anisotropy imposed by the external field while the other does not display any preferential direction.

The first set of initial conditions (anisotropic initial condition) complied with the anisotropy imposed by the external magnetic field. This initial condition was generated using a two-step process. First, a set of Fourier modes for both magnetic and velocity field fluctuations is produced, with amplitudes such that the (omnidirectional) energy spectrum is a Kolmogorov spectrum proportional to $k^{-5/3}$ for $1 \le k \le 16$ and gaussian random phases. Modes outside this range in k-space are set to 0. Second, an anisotropic filter is applied, so that excited modes for which $\tau_{nl}({\vk}) > \tau_A ({\vk})$ are initially set to 0 (Here $\tau_{nl}({\vk}) = 1/(k v_k)$ is the non-linear time associated with the speed $v_k$ at wavenumber $k$ and $\tau_A ({\vk}) = 1/(\kpar V_{A_0})$ is the Alfv\'en time, where $V_{A_0}=B_0/\sqrt{4\pi\rho_0}$ is the Alfv\'en speed). This is to ensure that only modes which satisfy the RHMHD requirement are excited initially \cite{rhmhd_2,montgomery,kado,i11}. Taking the Kolmogorov inertial range form $v_k \sim k^{-1/3}$ into account, this condition becomes
\begin{equation}
\kpar \le C \frac{k^{2/3}}{V_{A_0}}
\label{eq:RMHD_cond}
\end{equation}
where $C$ is a O(1) constant. Here we view this inequality as one of the conditions for validity of the RMHD approximation. For marginal attainment of the time scale inequality, we consider runs with $C=1$. The condition means that only large parallel wavelength modes (low $\kpar$) are allowed in the initial fluctuations, and becomes increasingly restrictive as the value of the mean magnetic field (i.e. $V_{A_0}$) is increased.

To complete the specification of the initial conditions, the fluctuations are normalized so that the initial mean square values of the magnetic and velocity field are both equal to 1 (unit value). Cross helicity and magnetic helicity are initialized at very small values (remind that if the normalized cross helicity is unity there is no turbulent dynamics). Only plane-polarized fluctuations are considered in this case, that is, fluctuations are perpendicular to the mean magnetic field, e.g., ${\bf v}({\bf x}, t=0) = {\bf v}_\perp({\bf x}, 0)$.

The runs performed throughout this paper do not contain any magnetic or velocity stirring terms, so that
the CMHD and RHMHD systems evolve freely.

In the second case, for the generation of initial conditions (isotropic initial condition), we consider initial Fourier modes in a shell in k-space $1 \le k \le 2$ at low wavenumbers, with constant amplitudes and random phases. Unlike the previous case, no initial anisotropic filter is applied here, so the initial conditions are spectrally isotropic. As in the anisotropic case, only plane-polarized fluctuations (transverse to the mean magnetic field) are included, so these are (low- to high-frequency) Alfv\'en mode fluctuations and not magnetosonic modes.

This set of initial conditions represent a more general situation that may arise in an application, where no initial spectral anisotropy 
is imposed (for instance, they could represent better the effect of isotropic driving).

The organization of this section is as follows. Subsection \ref{subsection_a} provides a brief comparison of the RHMHD model and its compressible version. In subsection \ref{subsection_b} we present the numerical results for anisotropic initial conditions. Subsection \ref{subsection_c} shows the numerical results for isotropic initial conditions.

\subsection{Incompressible RHMHD Vs Compressible RHMHD} \label{subsection_a}

\begin{figure}

\includegraphics[width=10cm]{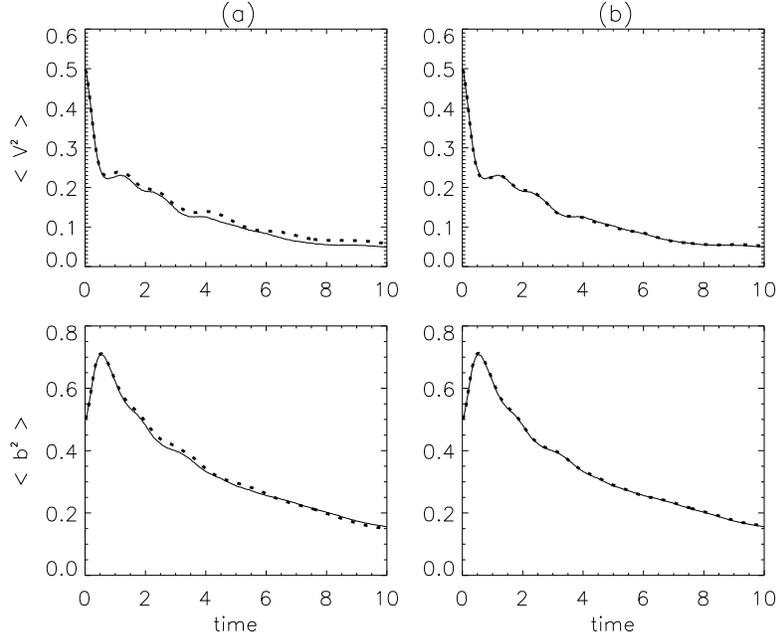}

\caption{\label{modelo_beta1}Kinetic energy (above) and magnetic energy (below) as function of time for the case $B_0=8$ for anisotropic initial conditions. Column ($a$) shows the RHMHD model and Column ($b$) the weak compressible RHMHD model. The solid lines corresponds to the CMHD model and the dotted lines to RHMHD models.}

\end{figure}

\begin{figure}

\includegraphics[width=10cm]{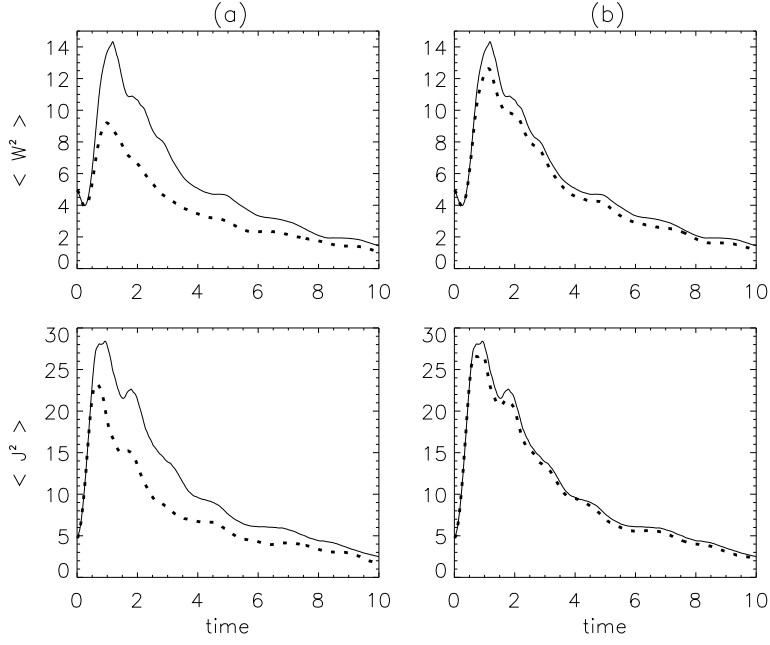}

\caption{\label{modelo_beta2}Vorticity (above) and current density (below) as function of time for the case $B_0=8$ for anisotropic initial conditions. Column ($a$) shows the RHMHD model and Column ($b$) the weak compressible RHMHD model. The solid lines correspond to the CMHD model and the dotted lines to RHMHD models.}

\end{figure}

With the introduction of the RHMHD model \cite{D2} we continue the line of incompressible reduced models (now considering the Hall effect), however it is possible to keep, at very low order, the terms responsible of the flow compression. To do this, we slightly changed the original RHMHD equations. 
In this section we compare one model to the other, to see how important is the role of compressibility in these systems.
The analysis performed on the results arising from the different simulations (both anisotropic and isotropic initial conditions) show a better fit with the full CMHD model when the compressible version of RHMHD is used. Figures (\ref{modelo_beta1}) and (\ref{modelo_beta2}) show different global parameters from the RHMHD simulations and its compressible version compared with the same parameters from the CMHD simulation.
Here $B_0=8$, and the initial condition has the anisotropy introduced from the external magnetic field. It can be seen that there are slight differences for the incompressible RHMHD model whereas a near-perfect fit is observed for the weak compressible RHMHD version. The spectra also show noticeable improvements when the compressible version was used. When the external magnetic field is changed or using isotropic initial conditions, the weakly compressible RHMHD model still performs better. This result indicates on one hand that the relevant parameters of the MHD turbulence in strong magnetic fields are affected by slight compressibility effects, and on the other hand that the weakly compressible RHMHD seems to fit correctly the above mentioned compressibility effects.

It is necessary to emphasize that between the RHMHD set of equations and its compressible version, there is no difference in the computational cost. Therefore, hereafter we adopt the compressible version, and refer to it simply with the initials RHMHD. Below we compare this (weakly compressible) RHMHD model vs the CMHD model.

\subsection{Anisotropic initial conditions} \label{subsection_b}

\begin{figure}

\includegraphics[width=10cm]{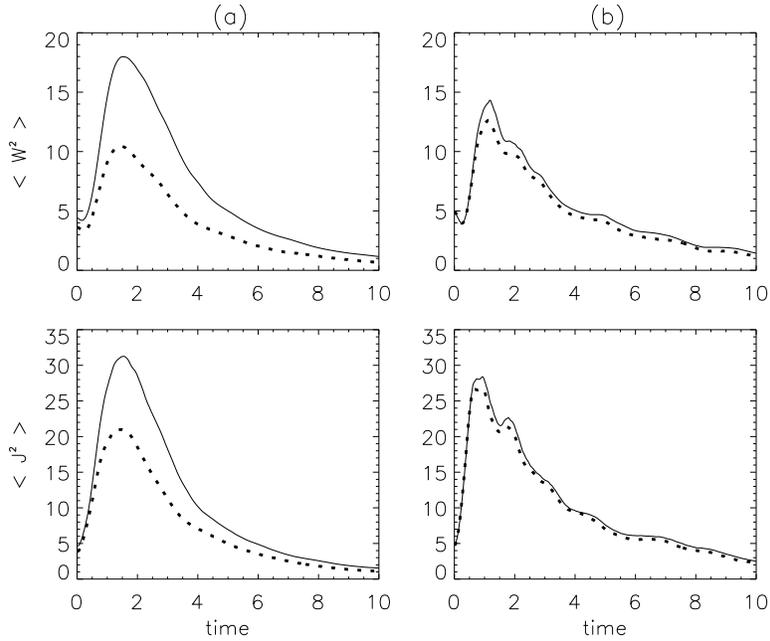}

\caption{\label{B1}Vorticity (above) and current density (below) as function of time for anisotropic case. Column ($a$): the results for $B_0=1$, the solid lines correspond to the CMHD model and the dotted lines to RHMHD model. Column ($b$): same for the results with $B_0=8$.}

\end{figure}

\begin{figure}

\includegraphics[width=10cm]{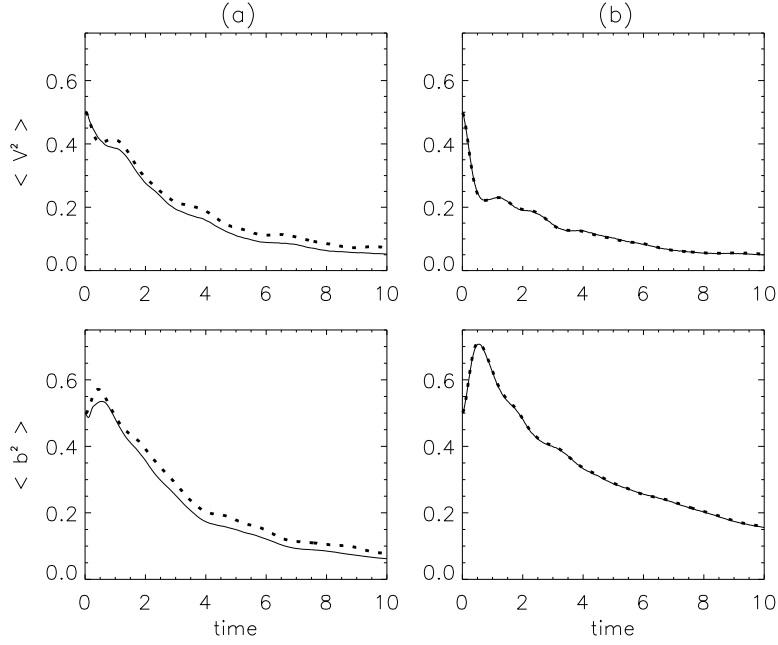}

\caption{\label{B2}Kinetic energy (above) and magnetic energy (below) as function of time for anisotropic case. Column ($a$): the results for $B_0=1$, the solid lines correspond to the CMHD model and the dotted lines to RHMHD model. Column ($b$): same for the results with $B_0=8$.}

\end{figure}

We carry out runs for two different external magnetic field intensities, $B_0=1$ and $B_0=8$. As expected, the global magnitudes studied show that the RHMHD equations fit better when the external field is sufficiently intense. Four of these magnitudes (enstrophy, current density, kinetic energy and magnetic energy) can be seen in figures \ref{B1} and \ref{B2}.

To obtain a more quantitative measure of the distance between the two solutions (RHMHD and CMHD), we calculate the average distance for the magnitudes shown in the figures  \ref{B1} and \ref{B2}. We define this parameter, for a given magnitude $a$, as
\begin{equation}
{\cal E}_a = \frac{\int (a ^{RHMHD} - a ^{CMHD})^2 ~~d^3t } {\int (a ^{CMHD})^2 ~~d^3t} 
\end{equation}
Table ~\ref{tabla1} shows this parameter for the magnitudes shown in Fig. \ref{B1} and \ref{B2}. Here we can see that when increasing the external field the average distance decrease three orders of magnitude for the magnetic energy, and two for the kinetic energy. The current density decreases two orders of magnitude and the enstrophy one order. There is more than 99,9 \% of agreement for kinetic and magnetic energy when $B_0=8$, and approximately 98 \% for enstrophy and  99,95 \% for current density. We can conclude that the RHMHD model is an optimal tool for characterize the global magnitudes when the initial condition satisfies the anisotropy imposed by the external field.

\begin{table}[H]

\caption{\label{tabla1} Anisotropic initial conditions}

\begin{ruledtabular}

\begin{tabular}{ccc}

Magnitude        & $B_0=1$    & $B_0=8$ \\ \hline

 ${\cal E}_{<{\bf V}^2>}$ & 0.0100301  &  0.000299110            \\ 

 ${\cal E}_{<{\bf B}^2>}$ & 0.0244613  &  $3.74040\times 10^{-5}$  \\ 

 ${\cal E}_{<{\bf W}^2>}$ & 0.189858   &  0.0112310              \\

 ${\cal E}_{<{\bf J}^2>}$ & 0.110552   &  0.00426486             \\ 

\end{tabular}

\end{ruledtabular}

\end{table}

\begin{figure}

\includegraphics[width=10cm]{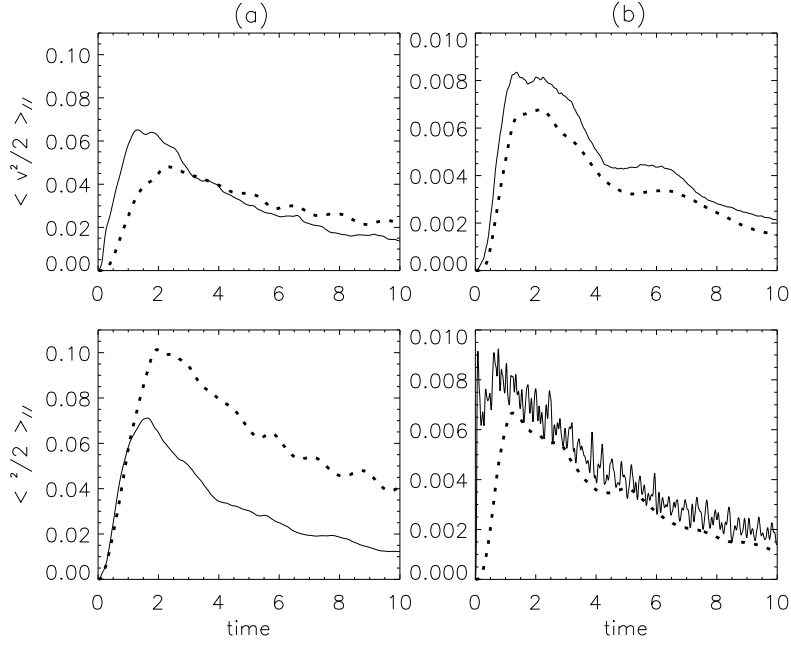}

\caption{\label{B3}Parallel component of the kinetic energy (above) and magnetic energy (below) as functions of time for anisotropic case. Column ($a$) the case $B_0=1$ and column ($b$) the case $B_0=8$. The solid lines correspond to the CMHD model and the dotted lines to RHMHD model.}

\end{figure}

In standard RMHD, the parallel components to the external magnetic field
do not affect the dynamics of the perpendicular fields, since
they do not appear in the dynamic equations for these fields
and in fact act as passive
scalars. If the parallel components are initially zero, they will remain
zero. This situation is, in principle, different in
RHMHD, because the Hall effect couples the dynamics of the perpendicular
components with the parallel components. This is evident in equations
(\ref{rhmhd1}) - (\ref{rhmhd4}). If the parallel components are initially
zero, they may not remain zero at subsequent times, because the Hall 
term act as a source in the evolution equations for those components.
Therefore, it is relevant to analyze the behavior of the parallel components when starting from initial conditions such that only perpendicular fields are not zero. Figure \ref{B3} shows the parallel velocity and magnetic field rms (root mean square) values.

\begin{figure}

\includegraphics[width=10cm]{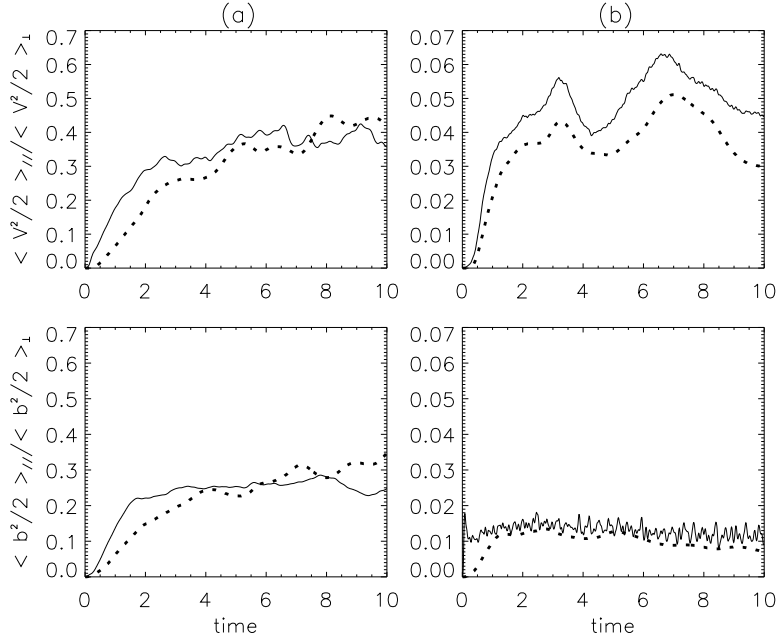}

\caption{\label{B4}Ratio between the parallel component and perpendicular component for kinetic (above) and magnetic energy (below). Column ($a$) the case $B_0=1$ and column $b$ the case $B_0=8$. The solid lines correspond to the CMHD model and the dotted lines to RHMHD model.}

\end{figure}

The ratio of parallel vs perpendicular components decrease with increasing external field, as shown in Fig.  \ref{B4}. In both cases the flow of energy between perpendicular components and the parallel component is stabilized. From the $B_0=1$ up to $B_0=8$ we can see that the parallel component goes from about 50 ~\% to less than 6 \% for the kinetic energy and 30 \% to 1.5 \% for the magnetic energy.

\begin{figure}

\includegraphics{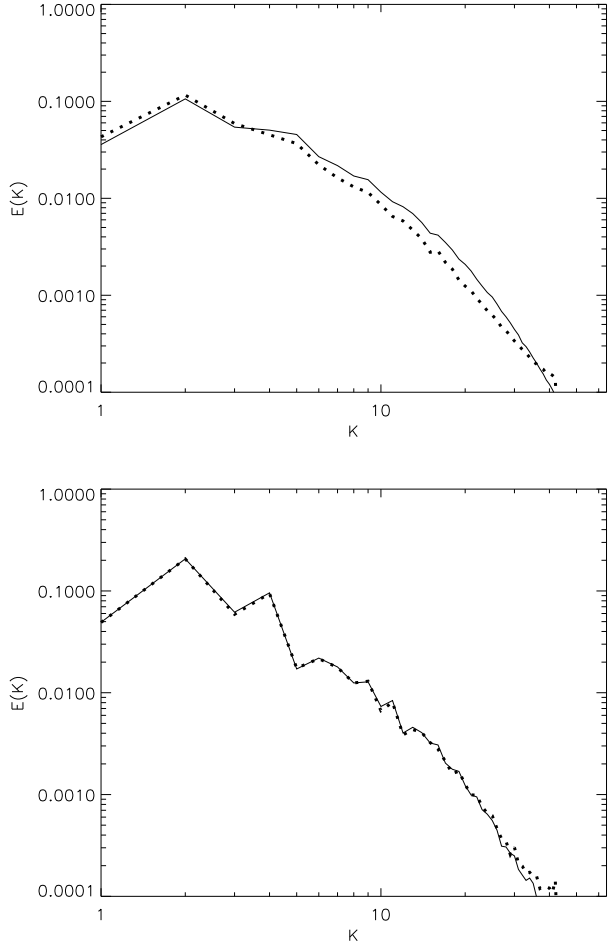}

\caption{\label{E1}Energy spectrum (kinetic plus magnetic) $E(k)$ for the cases $B_0=1$ (above) and $B_0=8$ (below), anisotropic initial conditions, and at $t=3$. The solid lines correspond to the CMHD model and the dotted lines to the RHMHD model.}

\end{figure}

Energy spectra for both $B_0=1$ and $B_0=8$ are shown in Fig. \ref{E1} at a time $t=3$. We can see that when the external field is small, there are slight differences between CMHD and the RHMHD model, but these differences disappear when the field increases. For the case $B_0=8$, the energy spectra of CMHD and RHMHD are identical. This shows that, when the external field is large enough and it maintains the anisotropy imposed by the initial conditions, the model reproduces the energy spectrum correctly. Therefore we can say that this is a robust model for spectral analysis in a anisotropic case.

\subsection{Isotropic initial conditions} \label{subsection_c}

In the last subsection we analyze the behavior of the solutions of the RHMHD equations whenever the initial condition satisfies the anisotropy imposed by the external field. We showed that the parallel fluctuations hardly contributed to the evolution of the system and therefore that its evolution is restricted only to the perpendicular plane to the magnetic external field. When we consider isotropic initial conditions (and therefore fluctuations along the parallel direction) new effects will appear, the most obvious of which are the Alfven waves. In this section we aim at  quantifying these new effects into the RHMHD model and see if this model continues to reproduce correctly the dynamics of the system. Beside, since only low-$k$ modes are excited initially, we expect an energy cascade to develop and populate the large wavenumber modes after a few turnover times. We also expect the nonlinear activity to be considerably stronger than in the anisotropic case.

\begin{figure}

\includegraphics[width=10cm]{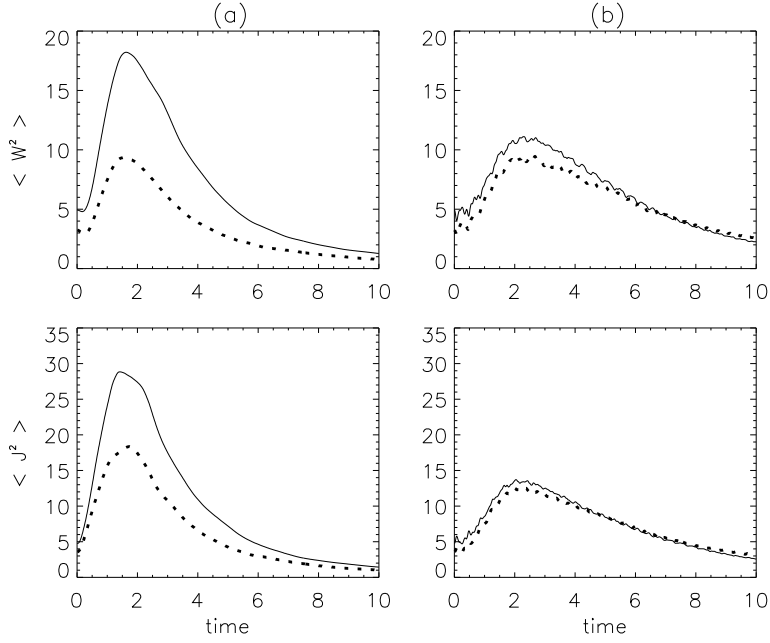}

\caption{\label{B1I}Vorticity (above) and current density (below) as function of time for isotropic case. Column ($a$): the results for $B_0=1$, the solid lines correspond to the CMHD model and the dotted lines to RHMHD model. Column ($b$): same for the results with $B_0=8$.}

\end{figure}

We perform simulations with $B_0=1$ and $B_0=8$, considering isotropic initial conditions. Figure \ref{B1I} shows the vorticity and current density as a function of time. Here we can see, again, a substantial improvement of the fit of the RHMHD curve to the CMHD curve when the external field becomes more intense. For $B_0= 8$ both the overall performance as well as the small-scale fluctuations are correctly
reproduced, the only difference that appears is a slight decrease in the amplitude of the RHMHD curve. However we again emphasize that the dynamics  is well captured for these global quantities.

\begin{figure}

\includegraphics[width=10cm]{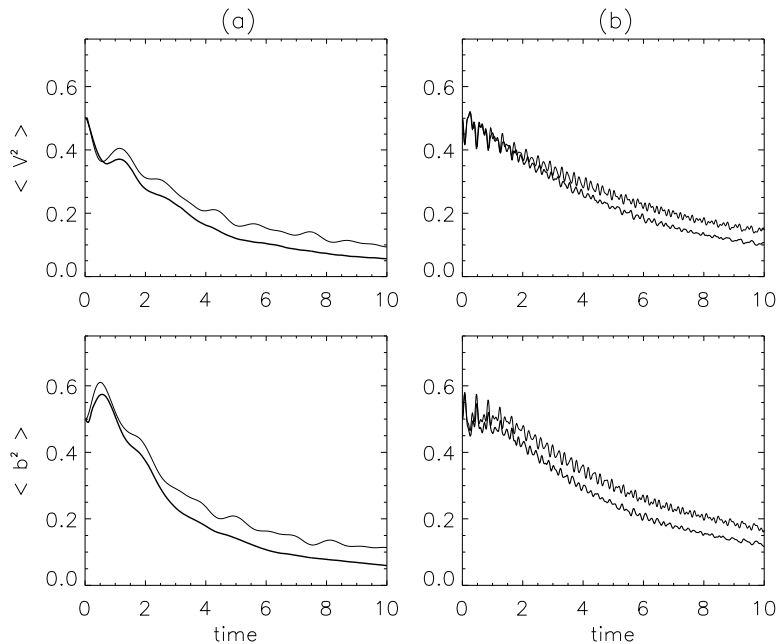}

\caption{\label{B2I}Kinetic energy (above) and magnetic energy (below) as function of time for isotropic case. Column ($a$): the results for $B_0=1$, the dark lines correspond to the CMHD model and the light lines to RHMHD model (we change the convention here so we can see Alfv\'en waves more clearly). Column ($b$): same for the results for $B_0=8$.}

\end{figure}

Figure \ref{B2I} shows the kinetic and magnetic energy as a function of time. Here we see that the perfect agreement obtained in the anisotropic case (when $B_0=8$) now gets lost. However the difference, as with the vorticity and current, is in an amplitude factor (now, a little bit more noticeable). Ignoring the differences due to the amplitude factor, we can see that the overall behavior is well described by the RHMHD model, in particular we can see that the fluctuations due to Alfven waves are correctly described.

Table ~\ref{tabla2} shows the average distance for global parameters for isotropic initial conditions. Here we can see that, although the average distances has increased in comparison to anisotropic initial conditions, these are maintained below 3 \%. Therefore the RHMHD model is still very suitable for the description of the system, when the external magnetic field is large.

\begin{table}[H]

\caption{\label{tabla2} Isotropic initial conditions}

\begin{ruledtabular}

\begin{tabular}{ccc}

Magnitude        & $B_0=1$    & $B_0=8$ \\ \hline

 ${\cal E}_{<{\bf V}^2>}$ & 0.0409073  &  0.0163489           \\ 

 ${\cal E}_{<{\bf B}^2>}$ & 0.0335319  &  0.0244613  \\ 

 ${\cal E}_{<{\bf W}^2>}$ & 0.249530   &  0.0193390              \\

 ${\cal E}_{<{\bf J}^2>}$ & 0.142143   &  0.00701709             \\ 

\end{tabular}

\end{ruledtabular}

\end{table}

\begin{figure}

\includegraphics{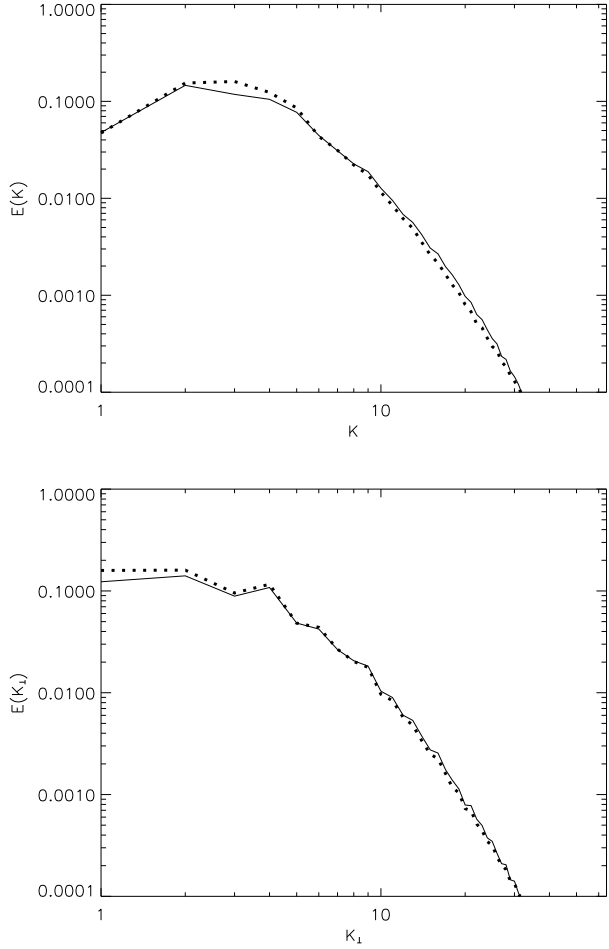}

\caption{\label{B3I}Energy spectrum (kinetic plus magnetic) $E(k)$ (above) and perpendicular energy spectrum $E(k_{\perp})$ (below) for the cases $B_0=8$ for isotropic initial conditions at $t=3$. The solid lines correspond to the CMHD model and the dotted lines to the RHMHD model.}

\end{figure}

In figure \ref{B3I}, we show the energy spectra (for $B_0=8$) at a time in which all the scales have been developed. Beside, we show the perpendicular energy spectra. Both spectra show a very good agreement of the RHMHD model with the full CMHD model.

The energy spectra for the RHMHD model shows a very good fit to the CMHD model, validating this model as a useful tool for spectral analysis of plasma embedded in strong magnetic fields even if the initial conditions do not maintain the anisotropy of the system. The reproducibility of the spectrum is a fundamental feature, because one of the open questions is whether the Hall effect might affect the energy inertial range. With this new model, which offers significant computational advantages, we expect to be able to address unsolved questions like this.

\section{Conclusions\label{conclusiones}}

We have numerically studied the validity of the RHMHD approximation
to the full equations of compressible 3D MHD (CMHD). As originally proposed 
in \cite{D3}, we took into account weak compressibility, without affecting the computational cost.

The weakly compressible RHMHD model quantitatively reproduces all 
the global quantities when a large external magnetic field is present. The quality of the fit with the 3D results show a slight decrease when the initial conditions do not comply with the anisotropy imposed by the external magnetic field. However, this decrease manifests itself in an amplitude factor, which does not alter significantly the dynamic evolution of the model.

We have also observed that parallel component fluctuations to the external
magnetic field remain small and do not grow with time.

The energy spectra are always well reproduced, regardless of whether the conditions are consistent with the symmetry of the system or not. Since important computational savings are
offered by the RHMHD model, we can conclude that it can be an excellent  
tool to study the spectral properties of the system, in particular the 
characterization of the inertial range and the dissipation range.

\acknowledgments
P. Dmitruk and D. O. G\'omez are members of the Carrera Investigador Cientifico CONICET. Research supported by grants X092/08, X429/08 from University of Buenos Aires, PICT 3370/05, PICT 00856/07 from ANPCyT and PIP 11220090100825 from CONICET.


\begin{thebibliography}{99}



\bibitem{D2}D. O. G\'omez, S. M. Mahajan, and P. Dmitruk, Phys. Plasma \textbf{15}, 102303 (2008).



\bibitem{i0} P. D. Mininni, D. O. G\'omez, and S. M. Mahajan, Astrophys. \textbf{J. 584}, 1120 (2003).



\bibitem{i1} F. Mozer, S. Bale, and T. D. Phan, Phys. Rev. Lett. \textbf{89}, 015002 (2002).



\bibitem{i2} D. Smith, S. Ghosh, P. Dmitruk, and W. H. Matthaeus, Geophys. Res. Lett. \textbf{31}, 02805, DOI: 10.1029/2003GL018689 (2004).



\bibitem{i3} L. F. Morales, S. Dasso, and D. O. G\'omez, J. Geophys. Res. \textbf{110}, A04204, DOI: 10.1029/2004JA010675 (2005).



\bibitem{i4} M. Wardle, Mon. Not. R. Astron. Soc. \textbf{303}, 239 (1999).



\bibitem{i5} S. A. Balbus and C. Terquem, Astrophys. \textbf{J. 552}, 235 (2001).



\bibitem{i6} W. H. Matthaeus, P. Dmitruk, D. Smith, S. Ghosh, and S. Ounghton, Geophys. Res. Lett. \textbf{30}, 2104, DOI: 10.29/2003GL017949 (2003).



\bibitem{i7} P. D. Mininni, D. O. G\'omez, and S. M. Mahajan, Astrophys. \textbf{J. 619}, 1019 (2005).



\bibitem{i8} S. Galtier, J. Plasma Phys. \textbf{72}, 721 (2006).



\bibitem{i9} P. Dmitruk and W. H. Matthaeus, Phys. Plasmas \textbf{13}, 2307 (2006).



\bibitem{rhmhd_2} H. R. Strauss, Phys. Fluids \textbf{19}, 134 (1976)



\bibitem{montgomery} D. C. Montgomery, Phys. Scr., T \textbf{T2/1}, 83 (1982) 



\bibitem{i11} G. P. Zank and W. H. Matthaeus, \textbf{J}, Plasma Phys. \textbf{48}, 85 (1992).



\bibitem{i12} A. A. van Ballegooijen, Astrophys. \textbf{J. 311}, 1001 (1986).



\bibitem{i13} D. W. Longcope, and R. N. Sudan, Astrophys. \textbf{J. 437}, 491 (1994).



\bibitem{i14} D. L. Hendrix and G. van Hoven, Astrophys. \textbf{J. 467}, 887 (1996).



\bibitem{i15} L. Milano, P. Dmitruk, C. H. Mandrini, D. O. G\'omez, and P. Demoulin, Astrophys. \textbf{J. 521}, 889 (1999).



\bibitem{i16} D. O. G\'omez, and C. Ferro Font\'an, Astrophys. \textbf{J. 394}, 662 (1992).



\bibitem{i17} P. Dmitruk and D. O. G\'omez, Astrophys. J. Lett. \textbf{527}, L63 (1999).



\bibitem{i18} P. Dmitruk, D. O. G\'omez, and W. H. Matthaeus, Phys. Plasmas \textbf{10}, 3584 (2003).



\bibitem{i19} S. Oughton, P. Dmitruk, and W. H. Matthaeus, Phys. Plasmas \textbf{11}, 2214 (2004).



\bibitem{i20} P. Dmitruk, W. H. Matthaeus, and S. Oughton, Phys. Plasmas \textbf{12}, 112304 (2005).



\bibitem{D3}N. H. Bian and D. Tsiklauri, Phys. Plasma \textbf{16}, 064503 (2009).



\bibitem{seudo} S. Ghosh, m. Hossain, and W. H. Matthaeus, Comput: Phys. Commun. \textbf{74}, 18 (1993)



\bibitem{kado} B. B. Kadomtsev and O. P. Pogutse, Sov. Phys. JETP \textbf{38}, 283 (1974)

 



\end{thebibliography}
\end{document}